\RequirePackage{fix-cm}
\documentclass[smallextended]{svjour3}       
\smartqed  
\usepackage{graphicx}
\usepackage{multirow}
\usepackage{hyperref}
\usepackage[table,xcdraw]{xcolor}
\usepackage{pdfpages}
\begin{document}

\title{Evaluating Software User Feedback Classifiers on Unseen Apps, Datasets, and Metadata}
\titlerunning{Evaluating Software User Feedback Classifiers on Unseen Data}        

\author{Peter Devine \and Yun Sing Koh \and Kelly Blincoe}


\institute{Peter Devine \and Kelly Blincoe \at
            Human Aspects of Software Engineering Lab,  University of Auckland, New Zealand \\
              \email{pdev438@aucklanduni.ac.nz; k.blincoe@auckland.ac.nz}           
           \and
           Yun Sing Koh \at
            School of Computer Science, University of Auckland
            }

\date{Received: date / Accepted: date}

\maketitle

\begin{abstract}

Listening to user's requirements is crucial to building and maintaining high quality software.
Online software user feedback has been shown to contain large amounts of information useful to requirements engineering (RE). Previous studies have created machine learning classifiers for parsing this feedback for development insight. While these classifiers report generally good performance when evaluated on a test set, questions remain as to how well they extend to unseen data in various forms.

This study evaluates machine learning classifiers performance on feedback for two common classification tasks (classifying bug reports and feature requests). Using seven datasets from prior research studies, we investigate the performance of classifiers when evaluated on feedback from different apps than those contained in the training set and when evaluated on completely different datasets (coming from different feedback platforms and/or labelled by different researchers). We also measure the difference in performance of using platform-specific metadata as a feature in classification.

We demonstrate that classification performance is similar on feedback from unseen apps compared to seen apps in the majority of cases tested. 
However, the classifiers do not perform well on unseen datasets. We show that multi-dataset training or zero shot classification approaches can somewhat mitigate this performance decrease.
Finally, we find that using metadata as features in classifying bug reports and feature requests does not lead to a statistically significant improvement in the majority of datasets tested.
We discuss the implications of these results on developing user feedback classification models to analyse and extract software requirements.

\end{abstract}

\section*{Conflict of interest}
None of the authors listed have a declared conflict of interest related to this work.

\maketitle

\section{Introduction}
    \label{sec:Introduction}
    Software product quality is deeply tied to user satisfaction and the extent to which the product meets the users' needs\cite{gillies2011software}. To that end, Requirements Engineering (RE) is considered key to developing high-quality software which meets users' needs~\cite{berki2004requirements,nuseibeh2000requirements}. 
    Recent research has found online software user feedback to be a rich source of information for understanding users' needs and the software requirements associated with those needs. For example, Pagano et al. showed that more than 30\% of 1100 manually analysed reviews on mobile app stores contained requirements relevant data that can then be leveraged by developers to improve their product~\cite{pagano2013user}. Similarly, feedback platforms such as Twitter posts~\cite{guzman2017little}, forum posts~\cite{tizard2019can}, Reddit posts~\cite{ali2020conceptualising}, Facebook posts~\cite{sultan2014back}, and Steam reviews~\cite{lin2019empirical}, have also been shown to contain helpful insights to guide the development and maintenance of software.
    
    Studies have proposed classification methods to automate the ingestion and analysis of this feedback to help identify software requirements~\cite{hadi2021evaluating,henao2021transfer,iacob2013online,panichella2016ardoc}. These methods are largely underpinned by machine learning models which require manually labelled example data to train on. Human annotators give labels such as ``bug report'' or ``feature request'' to each piece of feedback. These feedback-label pairs are then used as a training dataset to train a model to label feedback into one of these classes automatically. 
    
   The utility of these classifiers is multifaceted. 
   Proposals have been made for using classifiers to help developers understanding their users' requirements by integrating user feedback into the development cycle. 
  An example of this is MARA, which classifies app store reviews into ``feature request'' and ``bug report'', and these reviews are then used to inform software design and maintenance~\cite{iacob2013online}.
   Commercial solutions, such as MonkeyLearn\footnote{https://monkeylearn.com/}, also exist.
   In addition to aiding software teams to identify requirements, these classifiers can also be used in research studies. The classifiers can be trained on labelled training data and applied to large unlabelled datasets. This can enable researchers to study the requirements relevant characteristics of a large set of feedback, for example as was done by Nayebi et al. analysing user feedback on Twitter~\cite{nayebi2018app}.
    
    Despite classifiers widespread use throughout the literature, questions remain as to how effective these solutions are at classification out-of-the-box for software projects (i.e. without needing a representative sample of labelled feedback about the project as training data). Tizard et al. demonstrated that the popular ARDOC user feedback classifier, which has been trained on user feedback from app store reviews, does not perform well when applied to forum post feedback~\cite{tizard2019can}. Another concern is that many of these classification techniques rely on using feedback platform metadata (e.g., app review rating) as input features for classifiers - metadata which may not be available when applying such a classifier to another platform.
    Applying classifiers to feedback from unseen apps (i.e. apps of which none of its feedback was included in the classifier's training set) has also not been explicitly explored within the literature, which makes it unclear as to how they would perform on this unseen app feedback.
    If these classifiers are not able to correctly classify feedback from new apps and new platforms to a reasonably satisfactory degree, then they only have limited practical use in supporting the software development cycle.

    We investigate the robustness of user feedback classifiers over separate apps, domains, and features. To do this, we focus on three aspects of user feedback classifier training: training and testing on separate apps, training and testing on separate datasets, and training and testing both with and without metadata-based features.
    
    Firstly, many reported classification performance statistics are based on training and evaluating on user feedback from the same app, which leaves the expected performance of these classifiers on unseen apps unclear. 
    Therefore, this study examines the difference in classification performance between models trained and tested on feedback from separate apps, and trained and tested on the same apps. This is done to investigate how well models can classify feedback from unseen apps.
    
    Secondly, many public datasets of user feedback exist from prior studies. These datasets contain feedback from various feedback platforms (e.g. app store reviews, tweets, forum posts) and are labelled using various label sets. However, these datasets often contain labels that are similar as labels in the other datasets (e.g. ``bug'', ``bug report'', and ``error'' labels from three separate datasets). One labeller's definition of, for example, a bug report may differ from another's. Therefore, we evaluate whether a classifier trained on the labels of one dataset transfer to labelling another dataset. 
    This will provide an understanding on the ability of user feedback classification models to generalise to new domains or slightly different labelling schemas. 
    
    Thirdly, feedback metadata such as review rating and forum post position have also been used in many studies as a feature in classification. However, the effect of including metadata in classification on performance across multiple datasets is not fully known. Indeed, few studies examine the effect of each type of metadata to isolate the effect of metadata on overall performance. Therefore, we train classifiers both using and not using metadata as features to determine the change in performance with their inclusion. Understanding the relative importance of metadata on feedback classification informs how applicable they are to different sources which may have different metadata available.
    
    
    These aims resulted in the following three research questions:
    
    \begin{itemize}
        \item \textbf{RQ1:} How does training and testing on feedback from separate \textit{apps} affect classification performance?
        \item \textbf{RQ2:} How does training and testing on feedback from separate \textit{datasets} affect classification performance?
        \item \textbf{RQ3:} How does training and testing with \textit{metadata} affect classification performance?
    \end{itemize}
    
    We answer these questions by evaluating the classification performance of state-of-the-art text classifiers under different data configurations using seven datasets from the literature. We find little difference in classification performance between feedback from seen and unseen apps, but a large drop in performance on unseen datasets. We also find that using metadata as a feature in a classifier tends not to improve classification of bug reports and feature requests in the majority of cases.
    
    This paper first outlines the previous work related to user feedback classification in Section~\ref{sec:Related work}. The datasets used in this work are then detailed in Section~\ref{sec:Datasets} and the method used to train the classifiers is explained in Section~\ref{sec:Method}. The results of the evaluation on these datasets is described in Section~\ref{sec:Results}, and the implications of these results are described in Section~\ref{sec:Discussion}. Finally, threats to validity are considered in Section~\ref{sec:Threatstovalidity}.
    
    The replication package for this work can be found online\footnote{\href{https://doi.org/10.5281/zenodo.5733504}{https://doi.org/10.5281/zenodo.5733504}}.

\section{Related work}
    This section first details the effect that Requirements Engineering (RE) has on software quality, before exploring the applications of machine learning in RE within the literature. Finally, it gives examples of previous studies regarding the evaluation of different machine learning techniques in RE.

    \label{sec:Related work}
    
    \subsection{Software quality and requirements engineering}
        
        The benefits gained from RE has been shown to be integral to software quality. Many models exist within the literature that include RE to improve software quality~\cite{berki2004requirements,broy2006requirements}. 
        The practice of RE has also been shown to be beneficial, with Damian and Chisan demonstrating that productivity, quality, and risk management were all improved when effective RE was done within a commercial software project~\cite{damian2006empirical}.
        Similarly, Radliński showed that multiple RE factors had a positive, statistically significant effect on software quality factors within a literature dataset of thousands of software projects~\cite{radlinski2012empirical}.
        A survey of developers also showed that teams that used RE approaches were much more likely to say that their product’s capabilities fit their customers’ needs well and that end users found their products easy to use than those which did not~\cite{kassab2014state}.
        Other RE-adjacent concepts such as requirements traceability have also  been shown to positively impact software quality~\cite{rempel2016preventing}.
        
        
    \subsection{Machine learning in requirements engineering}
        Machine learning has become a common tool used within the requirements engineering literature for supporting the creation of requirements. Approaches like the one proposed by Cleland-Huang et al.~\cite{cleland2007automated} have been made to integrate automated text classifiers into the requirements engineering process. The MARA model, developed by Iacob et al.~\cite{iacob2013online} focuses on developing requirements from the ingestion of online user feedback using such a text classifier.
        These classifiers have become prevalent throughout the literature, with a systematic literature review by Lim et al. showing that 38 out of 63 studies which did user feedback analysis based on manually labelled data used machine learning to classify this feedback~\cite{lim2021data}.
        One of the potential reasons for this popularity is the reported high classification performance of some of these machine learning models.
        
        Work from Maalej et al.~\cite{maalej2016automatic}, Panichella et al.~\cite{panichella2016ardoc}, and Stanik et al.~\cite{stanik2019classifying} all report bug report classification F1 scores of app reviews higher than 0.75, with Maalej et al. reporting as high as 0.9. However, Stanik et al. also report a bug report classification score of 0.59 in Tweets, with Nayebi et al. also reporting lower feature request classification F1 scores of 0.67~\cite{nayebi2018app}.
        These diverse values highlight that the expected classification performance can vary dramatically depending on data source, classification method, and evaluation method. This underscores the need to rigorously compare and standardise techniques for both training and evaluating classification models.

    \subsection{Comparisons of classification techniques}
        There are several studies within the literature that compare techniques used for classifying user feedback.
        Work by Aurajo et al. evaluated the performance of four classical machine learning classifiers on classifying user feedback from one dataset using both term frequency derived features and features from deep pre-trained language models, showing that deep pre-trained language models generate superior text embedding features compared to frequency-based features for classification~\cite{araujo2020bag}. Similarly, Henao et al. demonstrated the increase in performance in user feedback classification when using pre-trained language models over both classical models as well as other deep models~\cite{henao2021transfer}. Hadi and Fard proposed a study where the classification accuracy of pre-trained language models is compared against that of previously constructed classifiers from the literature as well as exploring the effect of self-supervised pre-training, binary classification, multi-class classification, and zero-shot settings on classification performance~\cite{hadi2021evaluating}. Dhinakaran et al. showed that models trained on training data that was chosen randomly were found to consistently underperform more sophisticated training data selection techniques, such as active learning~\cite{dhinakaran2018app}. Di Sorbo et al. investigated the correlation between app review rating and feedback type classifier prediction, finding that predictions of ``problem discovery" from the ARDOC classifier were negatively correlated with the app rating, whereas predictions of ``feature request" were uncorrelated~\cite{di2021investigating}. As can be seen, there has been extensive work evaluating which text-based features and machine learning models are best to use when classifying user feedback. Some work has also been done to improve the data-efficiency of training a classifier. What remains unclear is how different training and evaluation methods affect the evaluation result of these classifiers (particularly on out-of-domain data), and how features apart from text affect classification performance.
        
        
    This study adds to the literature by exploring the effect of several machine learning techniques to highlight where and when user feedback classifiers can and cannot be used in the real world. Firstly, evaluating on seen and unseen app reviews is evaluated, in order to determine how well user feedback classifiers perform in classifying feedback for an unseen app. Secondly, classifiers trained and tested on separate datasets are evaluated so as to determine how well classifiers can be applied to similar data. Finally, the use of metadata for classification across multiple domains is evaluated to determine its effect on user feedback performance.

\section{Datasets}
\label{sec:Datasets}

    To measure different training and evaluation techniques, seven unique datasets from six studies were used in our evaluation. The datasets studied vary in size, feedback label set, and domain, coming from app reviews, Twitter, and forum posts. This variance among datasets was chosen partly to evaluate how well these classifiers perform across different domains and different labellers. 
    
    From these seven datasets, it was found that all seven shared a ``bug report'' or similar class, and six shared a ``feature request'' or similar class. Therefore, comparison of classification across datasets was done on a binary basis for these two classes.
    This section describes each dataset, with Table~\ref{tab:dataset-details} summarizing and comparing the broad statistics of each dataset.
    
    \begin{table}[]
\begin{tabular}{|l|l|r|r|l|l|l|r|r|}
\hline
\textbf{ID} & \textbf{Domain} & \multicolumn{1}{l|}{\textbf{Size}} & \multicolumn{1}{l|}{\textbf{\begin{tabular}[c]{@{}l@{}}No. \\ app\end{tabular}}} & \textbf{\begin{tabular}[c]{@{}l@{}}Bug \\ label\\ (\%)\end{tabular}} & \textbf{\begin{tabular}[c]{@{}l@{}}Feature \\ label\\ (\%)\end{tabular}} & \textbf{\begin{tabular}[c]{@{}l@{}}Meta-\\ data\end{tabular}} & \multicolumn{1}{l|}{\textbf{\begin{tabular}[c]{@{}l@{}}Bug \\ F1\end{tabular}}} & \multicolumn{1}{l|}{\textbf{\begin{tabular}[c]{@{}l@{}}Feature \\ F1\end{tabular}}} \\ \hline
\textbf{A} & Reviews & 1,565 & 27 & \begin{tabular}[c]{@{}l@{}}Error \\ (30.2\%)\end{tabular} & None & None & NR & NA \\ \hline
\textbf{B} & Reviews & 4,385 & 7 & \begin{tabular}[c]{@{}l@{}}Bug \\ report \\ (22.6\%)\end{tabular} & \begin{tabular}[c]{@{}l@{}}User \\ request\\ (9.2\%)\end{tabular} & Rating & 0.81 & 0.51 \\ \hline
\textbf{C} & Reviews & 1,438 & 48 & \begin{tabular}[c]{@{}l@{}}Bug \\ (9.5\%)\end{tabular} & \begin{tabular}[c]{@{}l@{}}Feature\\ (12.4\%)\end{tabular} & Rating & 0.88 & 0.85 \\ \hline
\textbf{D} & Reviews & 2,986 & 705 & \begin{tabular}[c]{@{}l@{}}Bug \\ (25.5\%)\end{tabular} & \begin{tabular}[c]{@{}l@{}}Feature\\ (11.1\%)\end{tabular} & \begin{tabular}[c]{@{}l@{}}Rating, \\ App \\ category\end{tabular} & 0.9 & 0.72 \\ \hline
\textbf{E} & Reviews & 707 & 14 & \begin{tabular}[c]{@{}l@{}}Bug \\ (69.3\%)\end{tabular} & \begin{tabular}[c]{@{}l@{}}Feature\\ (22.9\%)\end{tabular} & None & NR & NR \\ \hline
\textbf{F} & Forums & 2,652 & 2 & \begin{tabular}[c]{@{}l@{}}Apparent\\ bug \\ (15.3\%)\end{tabular} & \begin{tabular}[c]{@{}l@{}}Feature \\ request\\ (4.4\%)\end{tabular} & \begin{tabular}[c]{@{}l@{}}Post \\ position, \\ Topic\end{tabular} & \begin{tabular}[c]{@{}r@{}}0.725 \\ to \\ 0.728 \end{tabular} & \begin{tabular}[c]{@{}r@{}}0.83\end{tabular} \\ \hline
\textbf{G} & Twitter & 3,907 & 10 & \begin{tabular}[c]{@{}l@{}}Bug \\ (27.1\%)\end{tabular} & \begin{tabular}[c]{@{}l@{}}Feature\\ (24.2\%)\end{tabular} & None & 0.78 & 0.66 \\ \hline
\end{tabular}
\caption{Details of the 6 datasets used in our evaluation. (NR - not reported)}
\label{tab:dataset-details}
\end{table}
    
    The datasets included in this analysis are taken from replication packages linked in:
    \begin{itemize}
    \item \textbf{Dataset A} from Ciurumelea et al.~\cite{ciurumelea2017analyzing}\footnote{This dataset contains feedback that is labelled as ``Error''. While the classification of this class of feedback is not reported on in the paper, we use this class as our bug report class.}
    \item \textbf{Dataset B} from Guzman et al.~\cite{guzman2015ensemble}
    \item \textbf{Dataset C} from Maalej et al.~\cite{maalej2016automatic}
    \item \textbf{Dataset D} from Scalabrino et al.~\cite{scalabrino2017listening}\footnote{\label{scalabrino_footnote}The replication package contains two datasets referring to research questions 1 and 3 from this study, of which the latter is a pre-filtered set of feedback (filtered to contain only requirements relevant feedback) used to measure clustering performance, rather than classification. Therefore, while no classification metrics are reported for this RQ3 dataset (Dataset E), we still use it for training and testing models.}
    \item \textbf{Dataset E} from Scalabrino et al.~\cite{scalabrino2017listening}\textsuperscript{\ref{scalabrino_footnote}}
    \item \textbf{Dataset F} from Tizard et al.~\cite{tizard2019can}
    \item \textbf{Dataset G} from Williams et al.~\cite{williams2017mining}
    \end{itemize}
    
    Each dataset consists of a set of publicly available user feedback which has been scraped from the internet, before being manually labelled by the researchers of their respective studies. The smallest of these datasets has 707 pieces of feedback, while the largest has 4,385. The datasets span three distinct user feedback domains: app store reviews, forum posts, and tweets. The maximum number of distinct apps within a dataset was 705, while the smallest was two.

\section{Method}
    \label{sec:Method}
    
    To answer our research questions, we created training and test sets applicable to each experiment, before using the training sets to train state of the art text classifiers. Finally, we evaluated these models on test sets to get performance scores for each experiment.
    
    
\begin{table}[]
\begin{tabular}{|c|l|l|l|}
\hline
\multicolumn{1}{|l|}{}             &                                                                             & \textbf{Trained on}                                                                                              & \textbf{Tested on}                                                                               \\ \hline
\multirow{2}{*}{\textbf{RQ1}}      & Single mixed                                                                & \begin{tabular}[c]{@{}l@{}}Dataset $\alpha$ mixed \\ train and validation sets\end{tabular}                      & \begin{tabular}[c]{@{}l@{}}Dataset $\alpha$\\ mixed test set\end{tabular}                        \\ \cline{2-4} 
                                   & Single separated                                                            & \begin{tabular}[c]{@{}l@{}}Dataset $\alpha$ separated \\ train and validation sets\end{tabular}                  & \begin{tabular}[c]{@{}l@{}}Dataset $\alpha$\\ separated test set\end{tabular}                    \\ \hline
\multirow{2}{*}{\textbf{RQ2}}      & Single out-of-dataset                                                       & \begin{tabular}[c]{@{}l@{}}Dataset $\alpha$ mixed\\ train and validation sets\end{tabular}                       & \begin{tabular}[c]{@{}l@{}}Dataset $\delta$ \\ mixed test set\end{tabular}                       \\ \cline{2-4} 
                                   & Leave one out                                                               & \begin{tabular}[c]{@{}l@{}}Dataset $\alpha$, $\beta$, $\gamma$\\ mixed train\\ and validation sets\end{tabular}  & \begin{tabular}[c]{@{}l@{}}Dataset $\delta$ \\ mixed test set\end{tabular}                       \\ \hline
\multicolumn{1}{|l|}{\textbf{RQ3}} & \begin{tabular}[c]{@{}l@{}}Single mixed \\ (text and metadata)\end{tabular} & \begin{tabular}[c]{@{}l@{}}Dataset $\alpha$ mixed\\ train and validation sets\\ (text and metadata)\end{tabular} & \begin{tabular}[c]{@{}l@{}}Dataset $\alpha$ \\ mixed test set\\ (text and metadata)\end{tabular} \\ \hline
\end{tabular}
\caption{Example permutations of datasets for RQ1, RQ2, and RQ3 for example datasets $\alpha$, $\beta$, $\gamma$, and $\delta$.}
\label{tab:data-permutations}
\end{table}
    
    \begin{figure}[h]
        \includegraphics[width=\textwidth, angle=0]{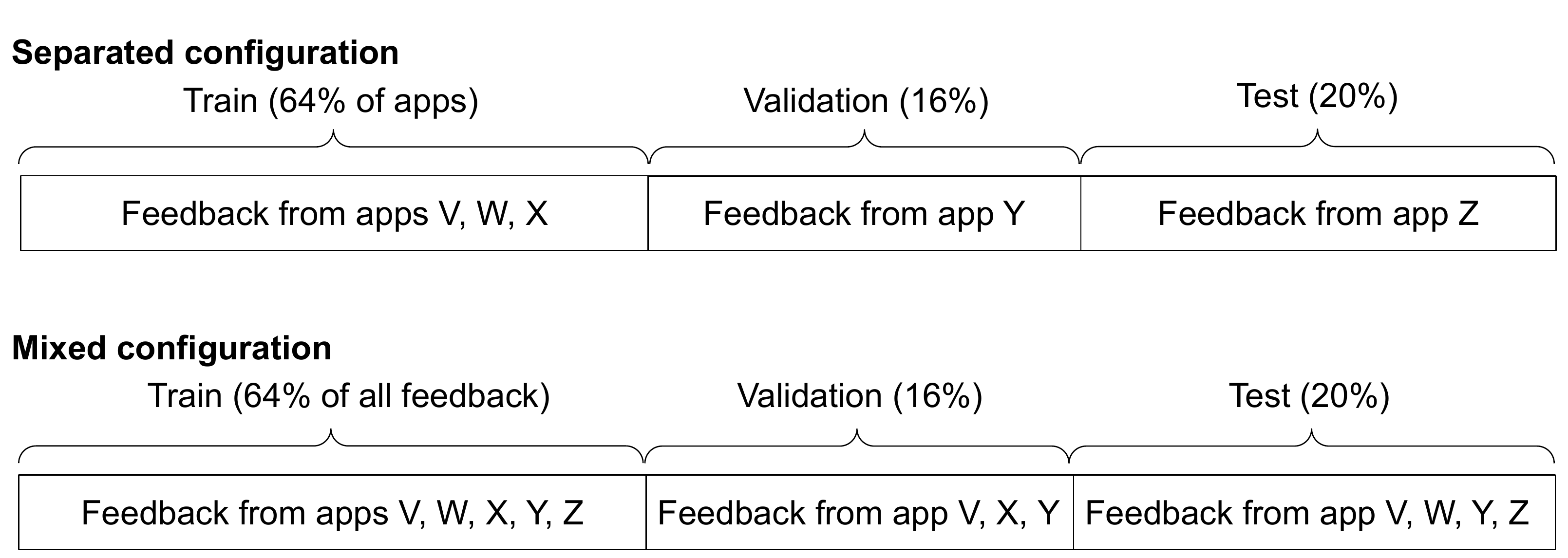}
        \caption{Diagram visualising how the train, validation, and test splits are created for both the mixed and separated datasets of RQ1.}
        \label{fig:RQ1-diagram}
    \end{figure}

    \subsection{Data handling}
        \label{subsec:Data handling}
        
        Within this study, the performance of classifiers on feedback from apps that it had not been trained on is analysed, and as such, the information as to which app a piece of feedback came from is needed for each dataset. For this reason each dataset was cleaned so that any rows which did not have an app identifier (i.e. are null) were dropped. This only affected dataset C, where some rows did not contain app identification data, and accounted for 1,821 out of 3,259 (55.9\%) rows within that dataset, leaving 1,438 pieces of feedback from that dataset to use in our experiments.
        
        Two different configurations of generating train, validation and test data splits were then used for each dataset, and these were named ``mixed'' and ``separated''.
        The ``mixed'' data splits were created by randomly splitting all feedback across the dataset into a 64:16:20 (train : validation : test) split. These ratios were chosen to be an initial 80:20 (train+validation : test) split before splitting the train + validation set with a further 80:20 split. This random splitting of data into train, validation, and test sets is current standard practice throughout the user feedback classification literature.
        
        The ``separated'' data splits were generating by randomly sampling apps within a dataset and assigning them to either the train, validation, or test split. All feedback for a given app was put into the same data split. Again, we aimed for a 64:16:20 (train : validation : test) ratio while ensuring each of these splits contained data from different apps. ``Separated'' data splits were not created for Dataset F because it included only two apps, and so contained too few distinct apps to split into train, validation, and test sets.
        A visual depiction of how these two configurations were created can be seen in Fig.~\ref{fig:RQ1-diagram}. 

        The ``mixed'' and ``separated'' configurations were achieved by using Sci Kit Learn’s ShuffleSplit and GroupShuffleSplit respectively.
        Cross validation was used to generate 5 distinct data folds for each dataset, and reported metrics in our results are the mean over these 5 folds.

    \subsection{Model}
        \label{subsec:Model}

        \subsubsection{Training machine learning models}
        We trained machine learning models based on state-of-the-art pre-trained language models, which have been shown to achieve higher classification performance than other models~\cite{henao2021transfer}. These models require text input to be tokenized before they can be trained.
    
        \label{subsec:Training language models}
        \textbf{Tokenization}
        
        To train and evaluate deep pre-trained language model based classifiers, we first tokenized all feedback text. This was done using Huggingface's Tokenizers library in Python \footnote{\href{https://huggingface.co/docs/tokenizers}{https://huggingface.co/docs/tokenizers}} using the ``bert-base-uncased'' version of the ``BertTokenizer'' tokenizer. This results in the creation of input IDs, an attention mask, and token type IDs for every piece of feedback. Each piece of feedback is also prepended by a [CLS] token and appended by a [SEP] token to denote the start and end of a piece of text. These values are then fed into the model for training or inference. We supply metadata tags to the model in the form of special tokens. All possible metadata tokens (i.e. those contained in train, validation, and test splits) are passed to the tokenizer when it is initialized such that it doesn't tokenize these features and that all metadata tags are valid input IDs when running training and inference.
        
        \textbf{Model training}
        
        The inputs generated by the tokenizer are then passed to a BERT model connected to one layer of two linear output nodes which create a head from which binary classification can be determined. This is done using the ``distilbert-base-cased'' version of the ``BertForSequenceClassification'' model from Huggingface's Transformers\footnote{\href{https://huggingface.co/transformers}{https://huggingface.co/transformers}} library in Python. This model variant was chosen due to it's relative high performance on general natural language tasks compared to larger language models~\cite{sanh2019distilbert} and because a smaller model allowed for more reasonable training times for the high number of models created within the constraints of this study. The model takes output from the first position (the [CLS] token) of the BERT model and passes it to the output linear layer, as is the norm when training BERT models. Labels are generated from this model using a softmax layer on top of the output to get a probability distribution between the in and out class of each piece of feedback. The class with the highest probability is then set as the prediction for the model.
        
        Each model is trained for 500 steps with a batch size of 128 (128 pieces of feedback training the model at each step). Every time all feedback has been used to traing the model (i.e. each epoch) the model is evaluated on the validation set. The weights of the model at the epoch with the highest associated F1 score on the validation set were loaded after training and saved for use in evaluating on the test set. The choice of training for 500 steps was made as it was observed that both the smallest and largest datasets had safely peaked in validation set F1 score by that point.
        
        A Trainer object from Huggingface's Transformers was used to train the model, into which we set a training batch size of 32. All other hyperparameters were left as default for the Trainer (initial learning rate = 5e-05, weight decay = 0, adam beta 1 = 0.9, adam beta 2 = 0.999, adam epsilon = 1e-08) as there was little observed difference in performance when these were changed.
    
        \subsubsection{Zero shot classifier}
        
        A zero shot classifier model is a text classifier that does not require any training data before being used. In our work, we use the ``bart-large-mnli'' model developed by Facebook as our zero shot model due to its performance and popularity on the HuggingFace model portal\footnote{\url{https://huggingface.co/models?pipeline\_tag=zero-shot-classification}}.
        This model was not explicitly trained to classify user feedback, but has been designed to classify text without pre-training on class-labelled training data (hence ``zero-shot'') by leveraging the entailment prediction abilities of natural language inference models, as proposed by Yin et al.~\cite{yin2019benchmarking}. Therefore, this classifier relies on the textual content of the label as well as the text content of the feedback, and so requires model labels to categorise text into. We  used ``bug report'' for classifying bug reports, and ``feature request'' for classifying feature requests.
        This classifier outputs a classification score between 0 and 1, rather than a simple label. Therefore, we take a class to have been detected if it has a score of greater than 0.5.

    \subsection{Evaluation metrics}
        \label{subsec:Evaluation metrics}
        Each dataset was evaluated using the F1 metric. The equation for this metric can be found in equations \ref{eq:precision}, \ref{eq:recall}, and \ref{eq:f1}. This metric provides a good measure of how well a class is being correctly labelled due to it balancing recall and precision, and is less sensitive to class imbalances in the data compared to the accuracy metric.
        \begin{equation}
            precision=\frac{No.\:True\:Positives}{No.\:True\:Positives+No.\:False\:Positive}
            \label{eq:precision}
        \end{equation}
        \begin{equation}
            recall=\frac{No.\:True\:Positives}{No.\:True\:Positives+No.\:False\:Negative}
            \label{eq:recall}
        \end{equation}
        \begin{equation}
            F1=2.\: \frac{precision\: .\: recall}{precision+recall}
            \label{eq:f1}
        \end{equation}
        
        Statistical significance between the performances of different classifier types were determined by using an independent two-sample t-test on the F1 metrics all folds of cross validation for one given test dataset.
    
    \subsection{Training and eval}
    
        \subsubsection{Unseen Apps (RQ1)}
        In order to determine the difference in performance between evaluating on feedback from unseen apps compared to seen apps, we trained models on the cross validation folds of each dataset's training and validation sets for both ``separated'' and ``mixed'' configurations, for evaluation on their respective test sets. In our results, we denote these models as \textbf{``Single dataset - separated''} and \textbf{``Single dataset - mixed''}. 

        \subsubsection{Unseen datasets (RQ2)}
        \begin{figure}[b]
            \includegraphics[width=\textwidth, angle=0]{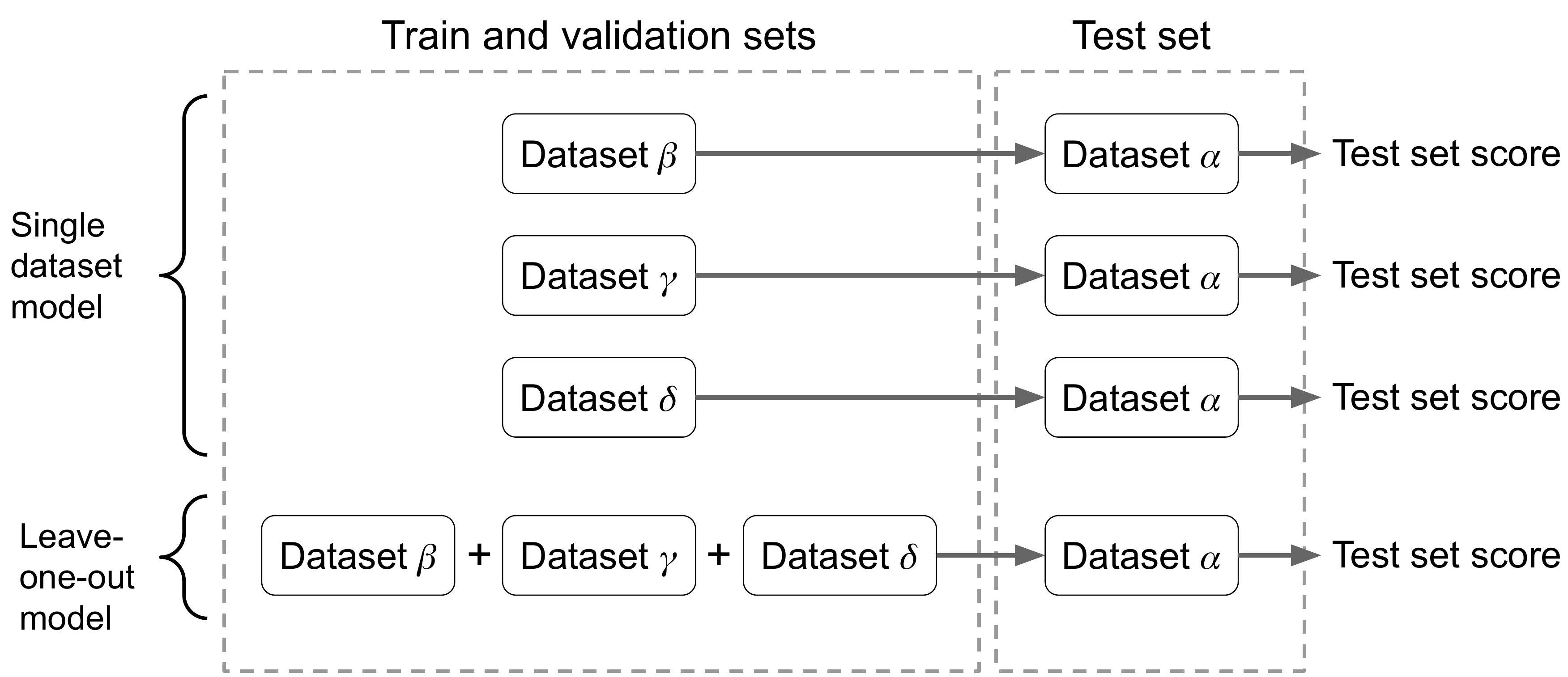}
            \caption{Diagram visualising how the training and testing is done on different datasets in RQ2.}
            \label{fig:RQ2-diagram}
        \end{figure}
        To find the classification ability of classifiers trained on one dataset before being applied to another, we used the models trained on ``Single dataset - mixed'' splits from RQ1 as the literature standard is to evaluate classifiers on mixed-app dataset splits. These were then evaluated on each dataset except the one that it was trained on. In our results, \textbf{``Train A-F''} denotes the models which have been trained on one of these datasets and is then evaluated on all others. In addition, we also trained a ``leave-one-out'' (denoted \textbf{``LOO''}) model for each data split, where all datasets except one were used to train a model, and then evaluated on the excluded dataset. A visual representation of how this training was done can be found in Fig.~\ref{fig:RQ2-diagram}. 

        For further context, a zero-shot text classification model (denoted \textbf{``Zero shot''}), as was proposed by Hadi and Fard~\cite{hadi2021evaluating}, was also evaluated on each dataset to provide a performance benchmark.
        
        \subsubsection{Metadata (RQ3)}
        \begin{figure}[t]
            \includegraphics[width=\textwidth, angle=0]{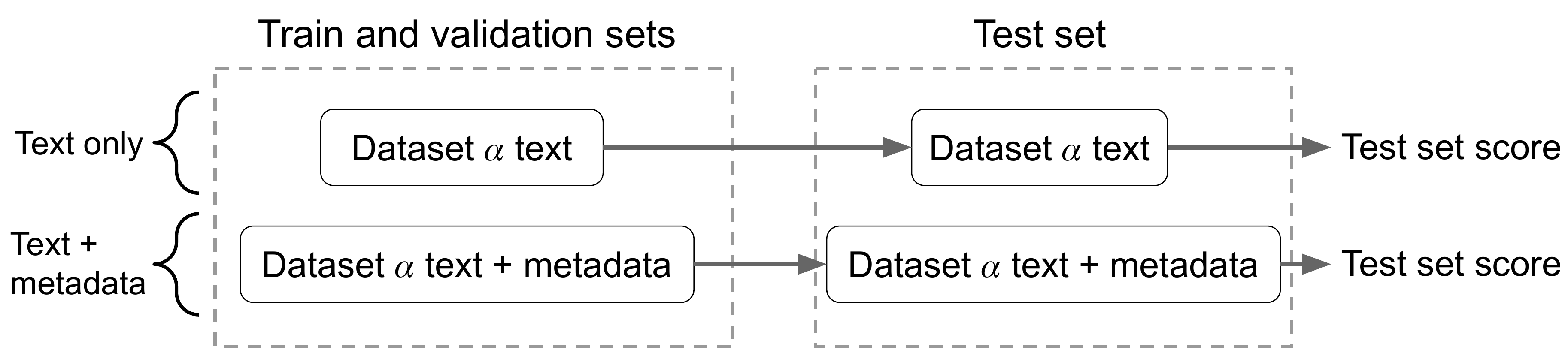}
            \caption{Diagram visualising how the training and testing is done using text, and text with metadata for datasets in RQ3.} \label{fig:RQ3-diagram}
        \end{figure}

        To determine the difference in performance of classifiers that use both metadata and text against those which use only text, we train two different models for every fold of every dataset, one which just receives feedback text as input, and one which also receives feedback metadata. As in RQ2, the ``Single dataset - mixed'' splits of data were used for these evaluations. A visual representation of this training and evaluation can be found in Fig.~\ref{fig:RQ3-diagram}.
        
        Metadata was added as a feature to the model by prepended feedback text with metadata tags before being passed to the model. These metadata tags are generated using the format as specified in Equation 1, such that an app review that has an associated rating of 3 stars would be prepended with the tag ``[METADATA\_rating\_3]''.
        
        \begin{equation}
            \texttt{[METADATA\_ + metadata column name + \_ + metadata value + ]}
            \label{eq:metadata format}
        \end{equation}
        
        Each metadata tag is added to the text tokenizer as a special token so as to prevent it from being broken up upon
        tokenization. We trained models using all metadata available to us from their datasets. This includes metadata that was used as features when making classifiers in the original studies associated with these datasets (e.g. follower count in dataset G).
        Full details of all metadata used with each dataset can be found in Table~\ref{tab:data-metadata}.
        After training on a given train and validation set, each model was evaluated on the respective test set.
        \textbf{``Text only''} denotes the evaluation results of the model which was trained and tested using only text features. \textbf{``Text and metadata''} denotes the evaluation results of the model which was trained and tested using both text and metadata features.
        
        \begin{table}[]
\begin{tabular}{|l|l|}
\hline
\textbf{Dataset} & \textbf{Metadata}                                                                                                                                                              \\ \hline
A                & App rating                                                                                                                                                                     \\ \hline
B                & App rating                                                                                                                                                                     \\ \hline
C                & App rating                                                                                                                                                                     \\ \hline
D                & \begin{tabular}[c]{@{}l@{}}App rating,\\ App category\end{tabular}                                                                                                             \\ \hline
E                & App rating                                                                                                                                                                     \\ \hline
F                & \begin{tabular}[c]{@{}l@{}}Post position,\\ Is comment author \\ original thread author,\\ Forum topic,\\ User level\end{tabular}                                              \\ \hline
G                & \begin{tabular}[c]{@{}l@{}}No. of favorites,\\ No. of followers,\\ No. of friends,\\ No. of statuses,\\ No of listed tweets\\ Is tweet a reply\\ Is user verified\end{tabular} \\ \hline
\end{tabular}
\caption{List of metadata used for datasets in RQ3}
\label{tab:data-metadata}
\end{table}

    A visual summary of how these three research questions were answered can be found in Table~\ref{tab:data-permutations}.
    
\section{Results}
    \label{sec:Results}
    This section first presents the results from training and testing on mixed and separated apps within datasets (RQ1). It then presents the results of training and testing on separate datasets (RQ2). Finally, it details the results of training and testing using metadata (RQ3).
    \subsection{Mixed vs. separated apps within data splits (RQ1)}
    Table~\ref{tab:rq1-all} details the mean F1 scores of classifiers from RQ1 for classifying bug reports and feature requests.

\begin{table}[]
\begin{tabular}{|l|rrrl|rrrl|}
\hline
 & \multicolumn{4}{c|}{\textbf{Bug classification}} & \multicolumn{4}{l|}{\textbf{Feature request classification}} \\ \cline{2-9} 
\textbf{Dataset} & \multicolumn{1}{l|}{\textbf{\begin{tabular}[c]{@{}l@{}}Single \\ dataset\\ separated\end{tabular}}} & \multicolumn{1}{l|}{\textbf{\begin{tabular}[c]{@{}l@{}}Single \\ dataset\\ mixed\end{tabular}}} & \multicolumn{2}{l|}{\textbf{\begin{tabular}[c]{@{}l@{}}T-test\\ stat.\end{tabular}}} & \multicolumn{1}{l|}{\textbf{\begin{tabular}[c]{@{}l@{}}Single \\ dataset\\ separated\end{tabular}}} & \multicolumn{1}{l|}{\textbf{\begin{tabular}[c]{@{}l@{}}Single \\ dataset\\ mixed\end{tabular}}} & \multicolumn{2}{l|}{\textbf{\begin{tabular}[c]{@{}l@{}}T-test\\ stat.\end{tabular}}} \\ \hline
\textbf{A} & \multicolumn{1}{r|}{0.664} & \multicolumn{1}{r|}{0.764} & -2.137 &  & \multicolumn{1}{r|}{-} & \multicolumn{1}{r|}{-} & - &  \\ \hline
\textbf{B} & \multicolumn{1}{r|}{\cellcolor[HTML]{EFEFEF}0.680} & \multicolumn{1}{r|}{\cellcolor[HTML]{EFEFEF}0.725} & -3.252 & * & \multicolumn{1}{r|}{0.399} & \multicolumn{1}{r|}{0.468} & -2.087 &  \\ \hline
\textbf{C} & \multicolumn{1}{r|}{0.339} & \multicolumn{1}{r|}{0.357} & -0.325 &  & \multicolumn{1}{r|}{0.501} & \multicolumn{1}{r|}{0.544} & -0.639 &  \\ \hline
\textbf{D} & \multicolumn{1}{r|}{0.868} & \multicolumn{1}{r|}{0.856} & 0.716 &  & \multicolumn{1}{r|}{0.659} & \multicolumn{1}{r|}{0.653} & 0.341 &  \\ \hline
\textbf{E} & \multicolumn{1}{r|}{\cellcolor[HTML]{EFEFEF}0.683} & \multicolumn{1}{r|}{\cellcolor[HTML]{EFEFEF}0.875} & -3.504 & ** & \multicolumn{1}{r|}{\cellcolor[HTML]{EFEFEF}0.443} & \multicolumn{1}{r|}{\cellcolor[HTML]{EFEFEF}0.642} & -2.580 & * \\ \hline
\textbf{G} & \multicolumn{1}{r|}{0.688} & \multicolumn{1}{r|}{0.704} & -0.837 &  & \multicolumn{1}{r|}{\cellcolor[HTML]{EFEFEF}0.528} & \multicolumn{1}{r|}{\cellcolor[HTML]{EFEFEF}0.597} & -3.227 & * \\ \hline
\end{tabular}
\caption{F1 score results for classifying bug reports and feature requests in RQ1 using both ``separated'' and ``mixed'' data splits. Student's independent t-test scores are also given with statistically significant differences shaded and highlighted with asterisks (* p\textless{}0.05, ** p\textless{}0.01, *** p\textless{}0.001).}
\label{tab:rq1-all}
\end{table}

    For the models which were trained on only one dataset, it can be seen that app-separated splits have a lower F1 score than mixed-app splits for 5 out of the 6 bug report datasets and 4 out of the 5 feature request datasets. For bug reports, two of these differences (B and E) were statistically significant to p\textless{}0.05, while for feature requests, two (E and G) were significant. Moreover, we see that only one of these differences is significant to p\textless{}0.01 over these 11 differences.
    
    
    \textbf{Answer to RQ1 - How does training and testing on feedback from separate apps affect classification performance?}
    In aggregate, a minority (4/11) of datasets exhibited statistically significant difference between being split along app lines compared to being split randomly.
    Training and testing a user feedback classifier on feedback from the same apps does not result in meaningfully different performance compared to training and testing on feedback from separate apps in a majority of cases.
    
    \subsection{Testing on separate datasets (RQ2)}
    Table~\ref{tab:rq2-bug} and Table~\ref{tab:rq2-feature} details the mean F1 scores of separate-dataset classifiers from RQ2 for classifying bug reports and feature requests, respectively. 

    \begin{table}[]
\begin{tabular}{|l|l|l|l|l|l|l|l|}
\hline
 & \textbf{A} & \textbf{B} & \textbf{C} & \textbf{D} & \textbf{E} & \textbf{F} & \textbf{G} \\ \hline
\textbf{Train A} &  & 0.413 & 0.315 & 0.430 & 0.595 & 0.140 & 0.301 \\
\textbf{Train B} & 0.628 &  & 0.311 & 0.716 & 0.690 & 0.267 & 0.425 \\
\textbf{Train C} & 0.540 & 0.551 &  & 0.585 & 0.668 & 0.173 & 0.408 \\
\textbf{Train D} & 0.330 & 0.465 & 0.278 &  & 0.541 & 0.158 & 0.391 \\
\textbf{Train E} & 0.467 & 0.364 & 0.185 & 0.399 &  & 0.203 & 0.439 \\
\textbf{Train F} & 0.191 & 0.177 & 0.108 & 0.282 & 0.161 &  & 0.212 \\
\textbf{Train G} & 0.645 & 0.581 & 0.324 & 0.624 & 0.817 & 0.317 &  \\ \hline
\textbf{Train LOO} & 0.694 & 0.704 & 0.422 & 0.781 & 0.792 & 0.344 & 0.493 \\ \hline
\textbf{Zero shot} & 0.653 & 0.645 & 0.360 & 0.712 & 0.743 & 0.394 & 0.692 \\ \hline
\textbf{Single dataset - mixed} & 0.764 & 0.725 & 0.357 & 0.856 & 0.875 & 0.455 & 0.704 \\ \hline
\end{tabular}
\caption{F1 scores for classifying bug reports in RQ2. Details the performance of being trained on one dataset and tested on another, as well as the performance of the ``leave-one-out'' (LOO) model, the zero-shot model, and the model trained on the same dataset as it is tested on. Note that the ``Single dataset - mixed'' model has been trained on in-domain data (i.e. the training set associated with the test dataset).}
\label{tab:rq2-bug}
\end{table}
    \begin{table}[]
\begin{tabular}{|l|l|l|l|l|l|l|}
\hline
 & \textbf{B} & \textbf{C} & \textbf{D} & \textbf{E} & \textbf{F} & \textbf{G} \\ \hline
\textbf{Train B} &  & 0.147 & 0.450 & 0.439 & 0.156 & 0.213 \\
\textbf{Train C} & 0.138 &  & 0.159 & 0.148 & 0.056 & 0.237 \\
\textbf{Train D} & 0.314 & 0.154 &  & 0.211 & 0.119 & 0.176 \\
\textbf{Train E} & 0.000 & 0.000 & 0.000 &  & 0.000 & 0.000 \\
\textbf{Train F} & 0.055 & 0.007 & 0.028 & 0.036 &  & 0.021 \\
\textbf{Train G} & 0.314 & 0.213 & 0.356 & 0.356 & 0.127 &  \\ \hline
\textbf{Train LOO} & 0.406 & 0.281 & 0.527 & 0.445 & 0.148 & 0.274 \\ \hline
\textbf{Zero shot} & 0.385 & 0.296 & 0.365 & 0.479 & 0.153 & 0.522 \\ \hline
\textbf{Single dataset - mixed} & 0.468 & 0.544 & 0.653 & 0.642 & 0.270 & 0.597 \\ \hline
\end{tabular}
\caption{F1 scores for classifying feature requests in RQ2. Details the performance of being trained on one dataset and tested on another, as well as the performance of the ``leave-one-out'' (LOO) model, the zero-shot model, and the model trained on the same dataset as it is tested on. Note that the ``Single dataset - mixed'' model has been trained on in-domain data (i.e. the training set associated with the test dataset).}
\label{tab:rq2-feature}
\end{table}

    As can be seen for both bug report and feature request classification, F1 score for any one given test dataset can vary wildly depending on the dataset of the training data. For bug report classifiers, we observe that the classifier trained only on dataset G training data (tweet user feedback) performs best on 4 of the 6 test datasets it is applied to. For feature request classifiers, the classifier trained only on dataset B (app review data) performs best on 4 of the 5 datasets it is applied to. The classifier trained in dataset F (forum data) performs worst on all test datasets it is applied to compared to other classifiers for bug reports. The classifier trained in dataset E (app reviews) performs worst on all test datasets for feature requests. Overall, every classifier trained on one dataset and evaluated on a separate dataset achieves lower classification performance compared to models trained and tested on the same dataset.
    
    In comparison to the models trained on only one (different) dataset, the leave-one-out classifier performs best on 6 of the 7 bug report datasets and 5 out of 6 feature report datasets that it is applied to. Compared to the model trained and tested on the same dataset, the leave-one-out classifier performs slightly worse on all but one dataset.
    
    The zero-shot classifier performs better than any of the single-dataset out-of-dataset for 5 out of the 7 bug report datasets and 4 out of the 6 feature request datasets. The zero shot model exceeds the performance of the leave-one-out models for the test sets of Dataset F and G (forum posts and tweets) for bug reports, and 4 out of 6 datasets (C, E, F, and G) for feature requests. Therefore, zero shot models perform best relative to other models on datasets of distinct feedback platforms.

    \textbf{Answer to RQ2 - How does training and testing on data from separate datasets affect classification performance?}
    Training and testing a user feedback classifier on feedback from separate datasets results in overall lower performance than training and testing on the same dataset. However, this lower performance can be improved upon by models trained on multiple datasets or by zero-shot text classification models.
    
    \subsection{Using metadata features to classify (RQ3)}
    Table~\ref{tab:rq3-all} details the mean F1 scores of RQ3 classifiers both including and excluding metadata features to classify feedback into bug reports and feature requests. 

\begin{table}[]
\begin{tabular}{|l|rrrl|rrrl|}
\hline
 & \multicolumn{4}{c|}{\textbf{Bug classification}} & \multicolumn{4}{l|}{\textbf{Feature request classification}} \\ \cline{2-9} 
\textbf{Dataset} & \multicolumn{1}{l|}{\textbf{\begin{tabular}[c]{@{}l@{}}Text \\ only\end{tabular}}} & \multicolumn{1}{l|}{\textbf{\begin{tabular}[c]{@{}l@{}}Text\\ and\\ Metadata\end{tabular}}} & \multicolumn{2}{l|}{\textbf{\begin{tabular}[c]{@{}l@{}}T-test\\ stat.\end{tabular}}} & \multicolumn{1}{l|}{\textbf{\begin{tabular}[c]{@{}l@{}}Text \\ only\end{tabular}}} & \multicolumn{1}{l|}{\textbf{\begin{tabular}[c]{@{}l@{}}Text\\ and\\ Metadata\end{tabular}}} & \multicolumn{2}{l|}{\textbf{\begin{tabular}[c]{@{}l@{}}T-test\\ stat.\end{tabular}}} \\ \hline
\textbf{A} & \multicolumn{1}{r|}{0.764} & \multicolumn{1}{r|}{0.817} & -2.059 &  & \multicolumn{1}{r|}{-} & \multicolumn{1}{r|}{-} & - &  \\ \hline
\textbf{B} & \multicolumn{1}{r|}{\cellcolor[HTML]{EFEFEF}0.725} & \multicolumn{1}{r|}{\cellcolor[HTML]{EFEFEF}0.761} & -3.270 & * & \multicolumn{1}{r|}{0.468} & \multicolumn{1}{r|}{0.475} & -0.168 &  \\ \hline
\textbf{C} & \multicolumn{1}{r|}{0.357} & \multicolumn{1}{r|}{0.432} & -1.481 &  & \multicolumn{1}{r|}{0.544} & \multicolumn{1}{r|}{0.535} & 0.171 &  \\ \hline
\textbf{D} & \multicolumn{1}{r|}{0.856} & \multicolumn{1}{r|}{0.857} & -0.155 &  & \multicolumn{1}{r|}{0.653} & \multicolumn{1}{r|}{0.669} & -0.698 &  \\ \hline
\textbf{E} & \multicolumn{1}{r|}{0.875} & \multicolumn{1}{r|}{0.878} & -0.172 &  & \multicolumn{1}{r|}{0.642} & \multicolumn{1}{r|}{0.687} & -0.791 &  \\ \hline
\textbf{F} & \multicolumn{1}{r|}{0.455} & \multicolumn{1}{r|}{0.522} & -1.985 &  & \multicolumn{1}{r|}{\cellcolor[HTML]{EFEFEF}0.270} & \multicolumn{1}{r|}{\cellcolor[HTML]{EFEFEF}0.463} & -5.102 & ** \\ \hline
\textbf{G} & \multicolumn{1}{r|}{0.704} & \multicolumn{1}{r|}{0.722} & -1.235 &  & \multicolumn{1}{r|}{0.597} & \multicolumn{1}{r|}{0.605} & -0.367 &  \\ \hline
\end{tabular}
\caption{F1 score results for classifying bug reports and feature requests in RQ3 both using and not using metadata based features. Student's independent t-test scores are also given with statistically significant differences shaded and highlighted with asterisks (* p\textless{}0.05, ** p\textless{}0.01, *** p\textless{}0.001).}
\label{tab:rq3-all}
\end{table}

    For bug reports, we find all datasets have higher F1 scores when metadata and text is used to classify compared to when only text is used. However, only one of these differences (dataset B) increases between models were statistically significant with a p-value of \textless{}0.05.
    
    Similarly for feature requests, we find that for 5 out of the 6 datasets studied, models which use metadata and text perform better than just using text. Again, only one (datasets F) of these differences were statistically significant to a p-value of 0.05.
    
    Overall, we can see either a slight increase or no change in the performance of classifiers when metadata and text are used together compared to when text alone.\\
   
    \textbf{Answer to RQ3 - How does training and testing with \textit{metadata} affect classification performance?}
    Training on metadata results does not result in a statistically significant increase in classification performance on the majority of datasets tested.
    
\section{Discussion}
    \label{sec:Discussion}
    This section discusses the results of this work and their implications. Firstly, the effect of training and testing on separate apps is discussed. Then the effect of training and testing on separate datasets is described. Finally, the effect of using metadata in user feedback classifiers is analysed.
    \subsection{RQ1 - Mixed vs. separated apps within data splits}
        From our results, we found little difference between evaluating a classification model on feedback from unseen apps compared to evaluating on feedback from the same apps that it was trained on. This finding is made across both models classifying  bug reports and feature requests. This result suggests that model evaluation as is currently carried out within the literature (i.e. not specifying that train, validation, and test splits must contain feedback from separate apps) can be seen to be a good predictor of performance of a classifier on unseen apps from within a dataset. This hints at potential real-world applicability of these models in that they could be used on feedback from unseen new apps (but crucially from the same platform and data-gathering process) without an expected drop in performance.
        
        Another outcome of these experiments is that the classification F1 score can range from high (greater than 0.8) to low (less than 0.5). When a classifier has low absolute classification performance, their utility in finding requirements is limited. This finding highlights the fact that automatic classification using current technology is not universally useful across all feedback datasets. The fact that many of these values are slightly lower than their literature quoted values could possibly be due to the fact that we decided against doing extensive hyperparameter tuning when training our classifiers. Reasoning and discussion of this is given in Section~\ref{sec:Threatstovalidity}.
        Finally, the lower classification performance across most of the datasets for classifying feature requests compared to bug reports is a trend that can be broadly seen throughout the literature, and calls into question exactly why a bug report is so much easier to identify (from a machine learning perspective) compared to a feature request.
        
        
    \subsection{RQ2 - Testing on unseen datasets}
        In RQ2, we found that a model trained on one dataset and then applied to another dataset achieves worse performance than a model trained and tested on the same dataset. This is not a surprising finding, given that class balance and labelling methods vary slightly between datasets. However, it raises an important question: How informative are the predictions of these models when used in the real world? A dataset of user feedback for a given software project is not guaranteed to have a certain class balance, and a given researcher or developer is not guaranteed to consider a piece of feedback to contain a bug or feature request in the same way that the training data labellers did.
        
        Our results with the leave-one-out models show better performance, in contrast. While the leave-one-out models perform worse than models trained and tested on the same data, they perform better and are more consistent compared to models trained on one dataset. The leave-one-out models also perform better compared to the zero shot classifier except for feedback from an unseen platform (tweets and forum posts) at training time. This indicates that while leave-one-out classifiers are useful, zero shot classifiers are more appropriate for classifying feedback from unseen feedback platforms.
        
        These results suggest that user feedback analysis tools will achieve highest performance if before use they first require a sample of labelled user feedback from the developer who intends to use the tool (i.e. use in-domain training data). This could be done through an active learning approach, such as was explored by Magalhães et al.~\cite{magal}, in order to limit the amount of labelled data required. However, a tool that requires no further labelled data before being used can still achieve good performance if it is either: trained using a labelled feedback dataset from the same feedback platform; or a zero-shot text classifier if no such dataset exists.
        
        With these results, we recommend that future creators of user feedback analysis tools train a classification model using as much labelled user feedback as possible, especially using data from the same platform as its intended use-case. If such a dataset does not exist and is prohibitively expensive to create, then we recommend using zero-shot classification models instead.
        
        In order to aid future user feedback analysis tools, we make bug report and feature request classifiers available for use on the Huggingface platform. We aim to make these available with a link upon publication.

    \subsection{RQ3 - Classifying with metadata}
        Our findings for RQ3 are that classification performance is modestly, but not significantly, improved when using metadata. This finding is in contrast to previous findings which reported the use of metadata on feedback classification, in which metadata was shown to have a positive impact on classification performance~\cite{maalej2016automatic,tizard2019can}. We theorise that this may be due to the fact that the state-of-the-art classifiers that we used contain millions of parameters~\cite{devlin-etal-2019-bert}, compared to very few parameters available in the classical machine learning models used in these earlier works. With this increased capacity, our model may be better able to infer metadata from the text itself (for example low review ratings would also be associated with more negative sentiment text), which means that having this information explicitly provided would not have much of an effect on the final prediction.
        We therefore recommend that the use of metadata as a feature should be reviewed within text-based software engineering machine learning tasks with the advent of new, very capacious language models such as BERT. Without the use of what appears to be largely superfluous metadata, these models are better able to be applied to different feedback and to new feedback platforms, where metadata may differ.
        
        
        We have shown that metadata does not affect performance significantly across a majority of datasets in the classification of bug reports and feature requests, but it is an open question as to how metadata would affect other classes of feedback. It is for future work to investigate a fuller picture as to which classes benefit most and least from use of metadata in their prediction.
        

\section{Threats to validity}
    \label{sec:Threatstovalidity}
 
    One threat to the validity of this work is that the results of this study many not generalise to the classification of other feedback classes. This study only examined the performance of classification models on classifying feedback into the binary labels of ``Bug report'' or ``No bug report'' and ``Feature request'' or ``No feature request''. These two labels were chosen due to the fact that they were the only two consistent labels across multiple datasets. Being able to automatically detect bug reports and feature requests from users is one of the key promises of utilizing online user feedback for requirements engineering~\cite{iacob2013online}. Furthermore, the abundance of these labels in various literature datasets highlights how useful these labels are considered to be. Therefore, focusing on the task of classifying bug reports and feature requests can still be seen to be valuable to those looking to engineer requirements using user feedback.
    It is for future work to replicate this research on other label types.
        
    Another potential threat to the validity of this work that we did not carry out any data balancing when creating our classifiers. Multiple studies within the literature, including those associated with datasets used in this work~\cite{maalej2016automatic,tizard2019can}, carried out data balancing before training their classifier. This is done to counteract the fact that user feedback may have classes of interest which are a small minority of overall feedback, and so a model is unable to learn the characteristics of this class if most of its training data is from other classes. However, studies on datasets outside of the domain of user feedback classification have shown that classifiers can perform well even when trained on highly unbalanced data~\cite{batista2004study}. Moreover, Henao et al. demonstrated that undersampling when training a deep language model has no major impact on the F1 score of the classifier~\cite{henao2021transfer}. It is for this reason that we decided against balancing our data, and it is for future work to fully explore the impact of data balancing on user feedback classification.
    
    A final threat to validity considered was the lack of hyperparameter tuning done for any one model, which may have led to lower absolute classification performance. While optimising the hyperparameters for any one app or dataset may have led to marginal performance gains, we found that in early experiments changing hyperparameters has little impact on overall classification performance.
    Our research questions also focused on the relative differences between machine learning treatments, rather than absolute values, and so we would expect that any performance improvements that would be introduced by hyperparameter tuning would not affect our overall conclusions.
    Furthermore, one of the aims of this work was to investigate how well models apply to unseen data domains. Tuning hyperparameters for the model's training domain may overfit it and disadvantage it when applied to out-of-domain data. It is for this reason we decided against extensive hyperparameter tuning.
    
\section{Conclusion}
    
    The technical quality of software is meaningless if it does not meet the needs of its intended users.
    Requirements engineering (RE) offers a way to gather the requirements of users, and has been shown to improve software quality generally.
    This work builds on the RE literature in understanding and automatically processing online user feedback for use in developing and maintaining software. Previous work has shown that it is possible to create text classifiers that can automatically detect bug reports, feature requests, and other requirements relevant information in user feedback for use in the software development cycle. This work contextualises these past results, and informs the future improvement of these classifiers. This has led to three broad contributions.
    
    Firstly, we showed that there can be a small drop in classification performance when applying trained classifiers to feedback from unseen apps for some datasets. However, this trend was found to be statistically insignificant across the majority of datasets tested. 
    
    Secondly, this paper demonstrated the classification performance of models which had not been trained on the dataset of given test set. We found that in the scenario where no data from a specific dataset is used to train a classifier, training a model on multiple other datasets achieves better performance than training on any one dataset alone. Moreover, we found that these multiple-dataset models are most applicable to datasets in which it contains feedback from platforms which the model has been trained on (app reviews). We found that for other platforms (tweets and forums), which did not have another dataset to represent it in the training data, zero-shot classification models performed better. 
    
    Finally, we demonstrate that  classification of both bug reports and feature requests do not notably benefit from metadata (app ratings, forum post position, etc.) as features. 
    
    Overall, these three results can inform the creation of better user feedback analysis tools so that, ultimately, developers will better understand the needs of their users and create higher quality software.
    
    We have made the replication package for this study available online\footnote{\href{https://doi.org/10.5281/zenodo.5733504}{https://doi.org/10.5281/zenodo.5733504}}.

    \bibliographystyle{spmpsci}      
\bibliography{sn-bibliography}
\end{document}